\documentclass{article}
\usepackage{amsmath,amssymb,amsthm}
\usepackage{enumerate}
\usepackage{hyperref}

\DeclareMathOperator{\conv}{conv}
\DeclareMathOperator{\aff}{aff}
\DeclareMathOperator*{\argmin}{arg\,min}

\DeclareMathOperator{\lex}{lex}
\DeclareMathOperator{\sspan}{span}

\newcommand{\R}{\mathbb{R}}
\renewcommand{\L}{\mathcal{L}}
\renewcommand{\epsilon}{\varepsilon}
\newcommand{\lexsucceq}{\succeq_{\lex}}
\newcommand{\tif}{\text{if }}
\renewcommand{\v}{\mathrm{v}}

\newtheorem{thm}[]{Theorem}
\newtheorem{proposition}{Proposition}
{\theoremstyle{definition}\newtheorem*{axiom}{Axiom}}
\newtheorem{alemma}[]{Lemma}
{\theoremstyle{definition}\newtheorem{example}[]{Example}}

\makeatletter
\newcommand{\axlabel}[2]{%
  \begingroup
    \protected@edef\@currentlabel{#1}
    \label{#2}%
  \endgroup
}
\makeatother

\usepackage[round]{natbib}
\title{Expected Utility Without Assuming Continuity}
\author{Gerrit Bauch\thanks{Center for Mathematical Economics, Bielefeld University, PO Box 10 01 31, 33501 Bielefeld, Germany. Email: gerrit.bauch@uni-bielefeld.de. The first draft circulated under the title ``Independence and indifferent points imply continuity''. For fruitful discussions, I thank Niels Boissonnet, Yves Breitmoser, Arthur Dolgopolov, Mikhail Freer, Simon Grant, Lorenz Hartmann, Edi Karni, Bernhard Kasberger, Jonathan Klinge, Christopher Kops, Tianyu Ma, Lasse Mononen, Fynn Närmann, Jurek Preker and Frank Riedel. The author gratefully acknowledges financial support by the German Research Foundation (DFG) [RTG 2865/1 – 492988838].}}
\date{\today}

\begin{document}

\maketitle

\begin{abstract}
    \noindent%
    I provide an axiomatization of expected utility in which topological continuity is replaced by a geometric axiom. The axiom requires a finite set of indifferent lotteries that span a hyperplane. In the case of three prizes, two indifferent lotteries suffice. The axiom is weaker than Solvability, as well as logically independent of Weak Continuity and the Archimedean axiom.
\end{abstract}

\section{Introduction}
Expected utility is classically derived from the Independence axiom and a topological continuity axiom. I show that continuity can be replaced by a finite geometric condition on indifference sets that I call Indifferent Points. Indifferent Points requires the existence of a finite set of indifferent lotteries whose affine hull spans a hyperplane; with three prizes, it reduces to indifference between two distinct lotteries. Under Independence, this condition is sufficient for -- and in fact equivalent to -- an expected utility representation. The result thus provides a behaviorally transparent and finitely testable alternative to standard continuity assumptions for expected utility preferences.

Indifferent Points is among the weakest notions for ruling out lexicographic orderings. In fact, I prove that Solvability, and thus also Strong Continuity and Mixture Continuity, imply Indifferent Points for weak orderings. By providing counterexamples, I show that Indifferent Points does not imply, and is not implied by, either the Archimedean Axiom or Weak Continuity.

Under Independence, \cite{hausner1954multidimensional} characterizes weak preferences on mixture spaces as lexicographic orderings of expected utility representations. From this perspective, Indifferent Points rules out lexicographic components by guaranteeing sufficiently many indifferent lotteries. As noted by \cite{fishburn1971study}, \citeauthor{hausner1954multidimensional}'s result may be less accessible for readers unfamiliar with ordered vector spaces; I therefore provide a direct geometric proof that Indifferent Points implies continuity and, hence, expected utility. \cite{ozbek2024expected} derives an expected utility representation by requiring the existence of a collection of pairs of indifferent lotteries, each satisfying a geometrically prescribed configuration. \citeauthor{ozbek2024expected}'s axiom system permits a slightly weaker form of Independence, albeit within a framework that assumes a unique worst lottery, which the key axioms invoke directly.

The article is organized as follows. Section \ref{Section: Framework and Analysis} contains the formal set-up, the new axiom and the expected utility representation. The axiom is logically compared to Solvability, Weak Continuity and the Archimedean axiom in Section \ref{Section: Relation}. The appendix contains the proofs.

\section{Framework and Analysis}\label{Section: Framework and Analysis}
Let $X = \{x_0, x_1, \ldots, x_n\}$ be a finite set of $n+1$ prizes where $n \geq 2$. Let $\L:= \Delta(X)$ be the set of lotteries over $X$, with generic elements $p,q,r$. I regard $\L$ as a mixture space by taking compound lotteries, i.e., $(\alpha p + (1-\alpha) q)(x_i) := \alpha p(x_i) + (1-\alpha) q(x_i)$ for $\alpha \in [0,1]$, $i \in \{0, \ldots,n\}$. Throughout, I consider a weak preference relation $\succeq$ on $\L$, i.e., $\succeq$ is a complete and transitive binary relation. As usual, $\sim$ and $\succ$ denote the symmetric and asymmetric parts of $\succeq$, indicating indifference and a strict preference. For fixed $p \in \L$, the strictly better, indifferent and strictly worse sets are defined as $\L_{\succ p} := \{ q \in \L \mid q \succ p \}$, $\L_{\sim p} := \{ q \in \L \mid q \sim p \}$, $\L_{\prec p} := \{ q \in \L \mid p \succ q \}$.

Classical expected utility theory features the Independence axiom, enforcing a form of linearity on preferences, and some form of continuity, ruling out lexicographic orderings, see, e.g., \cite{mas1995microeconomic}.

\begin{axiom}[Independence, IND]
    \axlabel{IND}{Axiom: IA}
    For all $p,q,r \in \L$ and $\alpha \in (0,1]$ we have
\begin{equation*}
    p \succeq q \iff \alpha p + (1-\alpha) r \succeq \alpha q + (1-\alpha) r.
\end{equation*}
\end{axiom}

\begin{axiom}[Mixture Continuity, MC]
    \axlabel{MC}{Axiom: MC}
    For all $p,q,r \in \L$, the sets $\{ \alpha \in [0,1] \mid \alpha p + (1-\alpha) q \succeq r \}$ and $\{ \alpha \in [0,1] \mid \alpha p + (1-\alpha) q \preceq r \}$ are closed in $[0,1]$.
\end{axiom}

Instead of topological continuity axioms like \ref{Axiom: MC}, I propose a strictly weaker, geometric axiom. This axiom requires the decision maker to be willing to make trade-offs, indicated by indifference among a finite set of linearly independent lotteries. To this end, regard $\L \subseteq \R^n$ as an $n$-dimensional simplex, identifying a lottery $p$ with the vector $ (p(x_1), \ldots, p(x_n)) \in \R^{n}$, i.e., leaving out the $0$-th coordinate which can be recovered by $p(x_0) = 1- \sum_{i=1}^n p(x_i)$.
The \emph{affine hull} of $p_1, \ldots, p_m \in \R^{n}$ is
\begin{align}
    \aff(p_1, \ldots, p_m) :&= \left.\left\{ \sum_{k=1}^m \lambda_k \cdot p_k \right| \sum_{k=1}^m \lambda_k = 1, \lambda_k \in \R \right\}\label{eq: affine hull sum = 1}\\
    &=\left.\left\{ p_1 + \sum_{k=2}^m \lambda_k \cdot (p_k-p_1) \right| \lambda_k \in \R \right\}.\label{eq: affine hull stutzvektor}
\end{align}
From expression \eqref{eq: affine hull sum = 1}, we see that the convex hull $\conv(p_1, \ldots, p_m)$ is contained in $\aff(p_1, \ldots, p_m)$. Expression \eqref{eq: affine hull stutzvektor} describes the affine hull as the translate of the vector space spanned by the direction vectors $\{p_k-p_1\}_{k=2}^m$ with $p_1$ as base point. The dimension $\dim \aff(p_1, \ldots, p_m)$ is defined as the $\R$-vector space dimension of $\sspan(\{ p_k-p_1 \mid k=2,\ldots,m \})$. An affine hull of dimension $n-1$ is a \emph{hyperplane}. The proposed axiom asserts the existence of a finite set of indifferent points that span a hyperplane.

\begin{axiom}[Indifferent Points, IP]
    \axlabel{IP}{Axiom: IP}
    There exist $p_1, \ldots, p_n \in \L$ with $p_1 \sim \ldots \sim p_{n}$ which span a hyperplane in $\R^n$.
\end{axiom}
In other words, the set of $n-1$ directional vectors $\{p_k -p_1\}_{k=2}^{n}$ is linearly independent in $\R^{n}$. Notably, in the case of $n=2$, often used to graphically illustrate expected utility theorems, \ref{Axiom: IP} is already fulfilled if the decision maker is indifferent between two distinct lotteries.

For a weak preference relation on $\L$ that fulfills \ref{Axiom: IA}, I show that continuity is equivalent to \ref{Axiom: IP} and therefore obtain an expected utility representation.

\begin{thm}\label{Theorem: main theorem}
    Let $\succeq$ be a weak preference on $\L$. Then the following statements are equivalent:
    \begin{enumerate}[(i)]
        \item $\succeq$ fulfills \ref{Axiom: IA} and \ref{Axiom: IP}.\label{item: THM IA and IP}
        \item $\succeq$ admits an expected utility representation, i.e., there exists $u \colon X \to \R$ such that $p \succeq q$ if and only if $\displaystyle\sum_{x \in X} p(x) \cdot u(x) \geq \sum_{x \in X} q(x) \cdot u(x)$.\label{item: THM EU}
    \end{enumerate}
\end{thm}

Since \ref{Axiom: IA} and \ref{Axiom: MC} imply expected utility, the interesting question is when a topological continuity axiom like \ref{Axiom: MC} can be replaced by \ref{Axiom: IP}, and conversely. In fact, Indifferent Points is implied by, and thus a weaker requirement than, many classical continuity notions like \ref{Axiom: MC}, even in the absence of the Independence axiom, cf.\ Section \ref{Section: Relation}. Under \ref{Axiom: IA}, continuity can be recovered from \ref{Axiom: IP} in through three observations: First, indifference between a set of lotteries carries over to all lotteries in their affine hull. Second, indifference is retained under translation. Finally, if the indifference class spans a hyperplane, it separates the simplex into two open half-spaces that are identified to be the strictly better and worse sets, therefore implying continuity.

\section{Relation to other Continuity Axioms}\label{Section: Relation}
Indifferent Points, \ref{Axiom: IP}, is a weaker axiom than many existing continuity axioms, such as Strong Continuity, Mixture Continuity, the Archimedean axiom, Weak Continuity or (weak Wold-)Solvability. The definitions can all be found in \citep{ozbek2024expected,ghosh2023continuity}. The following example shows that \ref{Axiom: IP} can hold while all continuity axioms mentioned above fail.

\begin{example}\label{Example: lexicographic with IP}
Let $\lexsucceq$ denote the lexicographic ordering on $\L \subseteq \R^n$, given by $p \lexsucceq q$ if and only if $p=q$ or there is $k \in \{1, \ldots, n\}$ such that $p(x_i) \geq  q(x_i)$ for all $i \in \{1, \ldots, k \}$ and $p(x_k) > q(x_k)$. Define a preference by writing $p \succeq q$ if and only if $p(x_1) = q(x_1) = \tfrac{1}{2}$ or $ p\lexsucceq q$. Then, $\succeq$ is complete, transitive and fulfills Indifferent Points \ref{Axiom: IP}, but is neither strongly continuous, mixture continuous, Archimedean, weakly continuous nor (weak Wold-)solvable. This can easily be seen by looking at a purely lexicographic domain, e.g., $p(x_1) > \tfrac{1}{2}$.
\end{example}

Furthermore, \ref{Axiom: IP} is indeed implied by many of the above mentioned continuity axioms and thus a strictly weaker assumption. In particular, we compare \ref{Axiom: IP} to the notion of Solvability.
\begin{axiom}[Solvability, Sol]
    \axlabel{Sol}{Axiom: SOL}
    For all $p,q,r$ with $p \succeq q \succeq r$, there exists an $\alpha \in [0,1]$ such that $\alpha p + (1-\alpha)r \sim q$.
\end{axiom}

\begin{proposition}\label{Proposition: Mixture Continuity implies IP}
    A weak preference satisfying Strong Continuity, Mixture Continuity or (weak Wold-)Solvability also fulfills Indifferent Points.
\end{proposition}

If $x_0$ is the unique worst (or best) lottery, \citeauthor{ozbek2024expected}'s geometric continuity axiom requires, for each face of the simplex having $x_0$ as a vertex, a pair of indifferent lotteries whose connecting line is parallel to a supporting hyperplane of that face.
\begin{axiom}[Weak Continuity, wCon]
    \axlabel{wCon}{Axiom: wCon}
    For all $i,j \in \{1, \ldots, n\}, i \neq j$, there exist $p,q \in \L$ with $p(x_k) = q(x_k) = 0$ for all $k \in \{1, \ldots, n\} \setminus \{i,j\}$, $p \sim q$ and $(p(x_i)-q(x_i)) \cdot (q(x_j) - p(x_j)) > 0$.
\end{axiom}

For $n=2$, \ref{Axiom: wCon} is a stronger assumption than \ref{Axiom: IP}, as it puts a restriction on the relative position of the indifference points. In general, however, neither condition implies the other: Example \ref{Example: lexicographic with IP} fulfills \ref{Axiom: IP}, but not \ref{Axiom: wCon}, while the example below satisfies \ref{Axiom: wCon}, but not \ref{Axiom: IP}.

\begin{example}
    Adapt a lexicographic ordering $\lexsucceq$ on $\L$ by making two appropriate lotteries on each relevant face indifferent. Concretely, define for each pair $i,j \in \{1, \ldots, n\}, i \neq j$ the lotteries $p^{i,j}(x_k) = \tfrac{1}{3}$, $q^{i,j}(x_k) = \tfrac{1}{4}$ if $k \in \{i,j\}$ and $p^{i,j}(x_k) = q^{i,j}(x_k) = 0$ for $k \in \{1, \ldots, n \} \setminus \{i,j\}$. Define $\phi \colon \L \to \L$, $\phi(p^{i,j}) = q^{i,j}$ and $\phi(p) = p$ if $p \neq p^{i,j}$ for any pair $i \neq j$ in $\{1, \ldots, n\}$. Setting $p \succeq q$ if and only if $\phi(p) \lexsucceq\phi(q)$ is thus a weak preference that satisfies \ref{Axiom: wCon} and violates \ref{Axiom: IP} for $n \geq 3$ as any indifference class has at most two elements.
\end{example}

Finally, I provide a counterexample that the Archimedean axiom does not imply \ref{Axiom: IP}. I first recap the axiom.
\begin{axiom}[Achimedean, Arch]
    \axlabel{Arch}{Axiom: Arch}
    For all $p,q,r \in \L$ with $p \succ q \succ r$, there are $\alpha, \beta \in (0,1)$ such that $\alpha p + (1-\alpha) r \succ q \succ \beta p + (1-\beta) r$.
\end{axiom}
By a result of \cite{karni2007archimedean}, \ref{Axiom: Arch} is equivalent to \ref{Axiom: MC} and local mixture dominance. The latter guarantees that, for any mixture of two lotteries, there exists a neighborhood of the scaling parameter in which small changes do not reverse the preference relation. Since \ref{Axiom: MC} implies \ref{Axiom: IP} by Proposition \ref{Proposition: Mixture Continuity implies IP}, this suggests that a counterexample is pathological. Indeed, I invoke the Axiom of Choice in the form of the well-ordering theorem and transfinite recursion in the construction.

\begin{example}
    In the following, I construct an injective map $\phi \colon \L \to [0,1]$ such that the image of every line segment in $\L$ is dense in $[0,1]$. Defining $p \succeq q$ if and only if $\phi(p) \geq \phi(q)$ then yields a strict preference relation that fulfills \ref{Axiom: Arch} and violates \ref{Axiom: IP}.\\
    To this end, let $\mathcal{S} = \{ \{ (\alpha p + (1-\alpha)q \mid \alpha \in [0,1] \} \mid p,q \in \L, p \neq q \}$ be the set of all non-degenerate line segments in $\L$ and $\mathcal{I} = \{ (a,b) \mid a,b \in \mathbb{Q} \cap [0,1], a < b \}$ be the set of all non-degenerate intervals in $[0,1]$ with rational endpoints. Let $\kappa$ be the initial ordinal of cardinality $\mathfrak{c}$ of the continuum. By the well-ordering theorem, we can well-order $\mathcal{S} \times \mathcal{I} = \{(S_{\xi}, I_{\xi}) \mid \xi \leq \kappa\}$ as well as $\L$ and $[0,1]$. In particular, any subset of $\L$ and $[0,1]$ has a minimal element w.r.t.\ the resp.\ well-order. Furthermore, let $E$ be a dense subset of $[0,1]$ with cardinality $\mathfrak{c}$ such that $F := [0,1] \setminus E$ also has cardinality $\mathfrak{c}$. Take $E$, e.g., to be the set of all numbers the first occurrence of the digit $7$ in its floating-point representation (if any) is in an even power of $1/10$.\\
    Now, for any ordinal number $\xi \leq \kappa$, construct $p_{\xi}\in S_{\xi}$ and $x_{\xi} \in I_{\xi} \cap E$ as follows. If $p_{\eta}, x_{\eta}$ are already constructed for all $\eta < \xi$, choose $p_{\xi} = \min S_{\xi} \setminus \{ p_{\eta} \mid \eta < \xi \}$ and $x_{\xi} = \min I_{\xi} \cap E \setminus \{ x_{\eta} \mid \eta < \xi \}$. These set differences are always non-empty as $\lvert S_{\xi}\rvert = \lvert I_{\xi} \cap E \rvert = \mathfrak{c}$. By the transfinite recursion theorem, we thus get an injective function $f \colon A := \{ p_{\xi} \mid \xi \leq \kappa \} \to E$, $f(p_{\xi}) = x_{\xi}$. Note in particular, that for any $S \in \mathcal{S}$, $x \in [0,1]$ and $\epsilon >0$, there is a $\xi$ such that $p_{\xi} \in S$ and $f(p_{\xi}) = x_{\xi}$ is $\epsilon$-close to $x$ by considering an $I \in \mathcal{I}$ with $I \subseteq (x-\epsilon, x + \epsilon) \cap [0,1]$. In other words, $f(A \cap S)$ is dense in $[0,1]$ for all $S \in \mathcal{S}$. To conclude the construction, choose any injection $g \colon \L \setminus A \to F$, which is possible since $\lvert \L \setminus A \rvert  \leq \lvert F \rvert = \mathfrak{c}$. Finally, define $\phi$ as $f$ on $A$ and $g$ on $\L \setminus A$.
\end{example}

\bibliography{ref}

@article{ozbek2024expected,
  title={Expected utility, independence, and continuity},
  author={Ozbek, Kemal},
  journal={Theory and Decision},
  volume={97},
  number={1},
  pages={1--22},
  year={2024},
  publisher={Springer}
}

@book{bourbaki,
  author    = {Bourbaki, Nicolas},
  title     = {Topologie g{\'e}n{\'e}rale. Chapitres 5 {\`a} 10},
  series    = {{\'E}l{\'e}ments de math{\'e}matique},
  publisher = {Springer},
  address   = {Berlin, Heidelberg},
  year      = {2007},
  note      = {Reprint of the original 1974 Hermann edition (Paris).},
  language  = {French}
}

@article{hausner1954multidimensional,
  title={Multidimensional utilities},
  author={Hausner, Melvin},
  journal={Decision processes},
  pages={167--180},
  year={1954},
  publisher={Wiley New York}
}

@article{karni2007archimedean,
  title={Archimedean and continuity},
  author={Karni, Edi},
  journal={Mathematical Social Sciences},
  volume={53},
  number={3},
  pages={332--334},
  year={2007},
  publisher={Elsevier}
}

@article{fishburn1971study,
  title={A study of lexicographic expected utility},
  author={Fishburn, Peter C},
  journal={Management Science},
  volume={17},
  number={11},
  pages={672--678},
  year={1971},
  publisher={INFORMS}
}

@article{ghosh2023continuity,
  title={Continuity postulates and solvability axioms in economic theory and in mathematical psychology: a consolidation of the theory of individual choice},
  author={Ghosh, Aniruddha and Khan, M Ali and Uyan{\i}k, Metin},
  journal={Theory and Decision},
  volume={94},
  number={2},
  pages={189--210},
  year={2023},
  publisher={Springer}
}

@book{mas1995microeconomic,
  title={Microeconomic theory},
  author={Mas-Colell, Andreu and Whinston, Michael Dennis and Green, Jerry R and others},
  volume={1},
  year={1995},
  publisher={Oxford university press New York}
}

\appendix
\section{Proofs}
\setcounter{alemma}{0}
\renewcommand{\thealemma}{\Alph{section}\arabic{alemma}}

I start with the proof of Proposition \ref{Proposition: Mixture Continuity implies IP} as it proves parts of the Theorem.
\begin{proof}[Proof of Proposition \ref{Proposition: Mixture Continuity implies IP}]
    I first prove that if there are at most countably many indifference classes, \ref{Axiom: IP} is always fulfilled. To this end, write $\L = \bigcup_{p \in \L}\L_{\sim p} = \bigcup_{p \in \L} (\aff(\L_{\sim p}) \cap \L)$, where each $\aff(\L_{\sim p}) \cap \L$ is a closed subset of $\L$. As a compact Hausdorff space, $\L$ is \emph{Baire space}, i.e., a countably union of closed sets contains an open set if and only if one of the involved closed sets does, cf.\ \cite{bourbaki}. If $\L$ only has countably many indifference classes, there is thus a $p \in \L$ with $\aff(\L_{\sim p}) \cap \L = \L$, implying, and even overachieving, \ref{Axiom: IP}.\\
    Assume thus that there are uncountably many indifference classes. I only consider Solvability \ref{Axiom: SOL} as it is the weakest among the continuity notions in question. Consequently, there are $p \succ q \succ r$ and we may assume that $q$ lies on the line segment joining $p$ and $r$ by \ref{Axiom: SOL}. We inductively construct a sequence $q_1, \ldots, q_{n}$ of indifferent points such that $\aff(q_1,\ldots,q_k)$ does not contain $p$ or $r$ and $\dim \aff(q_1,\ldots,q_k) = k-1$ for all $k \in \{1, \ldots, n\}$. \ref{Axiom: IP} then follows for $k=n$. Start with $q_1 := q$. Now, let $q_1, \ldots, q_k$ be already constructed and note that $\dim \aff(q_1, \ldots, q_k, p,r) = k$, since $p,q_1,r$ are collinear and $p,q \notin \aff(q_1, \ldots, q_k)$. As long as $k < n$, $\L \setminus \aff(q_1, \ldots, q_k, p,r)$ is non-empty and we can pick a point $s \in \L \setminus \aff(q_1, \ldots, q_k, p,r)$. If $s \sim q=q_1$, set $q_{k+1} := s$ such that $\aff(q_1,\ldots,q_{k+1})$ does not contain $p,r$ and has dimension $k$. If $s \succ q$, use \ref{Axiom: SOL} to find an $\alpha \in (0,1)$ with $q_{k+1} := \alpha s + (1-\alpha) r \sim q$. Again, $\aff(q_1,\ldots, q_{k+1})$ does not contain $p,r$ by construction and has dimension $k$. The case $s \prec q$ follows similarly by applying \ref{Axiom: SOL} to $p \succ q \succ s$.
\end{proof}

Before proving Theorem \ref{Theorem: main theorem}, I first provide some auxiliary lemmata. Except for the last one, these only require a weaker form of Independence, \emph{Betweenness}, that restricts \ref{Axiom: IA} to mixtures with $r \in \{p,q\}$.
\begin{axiom}[Betweenness, BET]
    \axlabel{BET}{Axiom: BET}
    For all $p,q \in \L$ and $\alpha \in (0,1)$, we have that $p \succ q$ (resp.\ $p \sim q$) implies 
        $p \succ \alpha p + (1-\alpha) q \succ q$ (resp.\ $p \sim \alpha p + (1-\alpha) q \sim q$).
\end{axiom}

\begin{alemma}\label{Lemma: affine space preserves indifference}
    Let $\succeq$ fulfill \ref{Axiom: BET} and let $p_1, \ldots, p_m \in \L$ with $p_1 \sim \ldots \sim p_m$. Then $\conv(p_1,\ldots,p_m)\subseteq
    \aff(p_1, \ldots, p_m) \cap \L \subseteq \L_{\sim p_1}$.
    \begin{proof}[Proof of Lemma \ref{Lemma: affine space preserves indifference}]
        Note that $\L_{\sim p}$ is convex, since if $q,q' \in \L_{\sim p}$, then also $\alpha q + (1-\alpha) q' \sim p$ by \ref{Axiom: BET} and transitivity. Consequently, $\conv(p_1, \ldots, p_m) \subseteq \L_{\sim p}$. Let $A := \aff(p_1, \ldots, p_m)$. If $m=1$, $A = A \cap \L = \{p_1\} \subseteq \L_{\sim p_1}$. Assume thus $m \geq 2$ in the following. By discarding linear dependent ones, we may assume that the vectors $p_k-p_1$ for $k=2, \ldots, m$ are linearly independent, making the representation of any element $p \in A$ in the form of expression \eqref{eq: affine hull sum = 1} unique. Let $p = \sum_{k=1}^m \lambda_k p_k \in \L$, $\sum_{k=1}^m \lambda_k =1$. If $\lambda_k \geq 0$ for all $k=1,\ldots, m$, we have $p \in \conv(p_1, \ldots, p_m)$ which we have already seen implies $p \in \L_{\sim p_1}$. Assume thus that $\lambda_k < 0$ for at least one $k \in \{ 1, \ldots, m\}$ and let $k^* \in \argmin_k \lambda_k$ and $\lambda^* := -\lambda_{k^*}$, which is positive. Set $\overline{p} := \sum_{k=1}^m \tfrac{1}{m} p_k \in \conv(p_1,\ldots,p_m) \subseteq \L_{\sim p_1}$. Define $q(\alpha) := \alpha \overline{p} + (1-\alpha) p$, which can be written as $q(\alpha) = \sum_{k=1}^m \left( \tfrac{\alpha}{m} + (1-\alpha)\lambda_k \right) \cdot p_k$. For $\alpha^* := \frac{m \lambda^*}{1 + m \lambda^*} \in (0,1)$, the coefficients of $q^*:=q(\alpha^*)$ are all non-negative and the one for $k^*$ is equal to zero. Thus, $q^* \in \conv(p_1,\ldots, p_m) \subseteq \L_{\sim p_1}$ and $q^* \neq \overline{p}$. By completeness, \ref{Axiom: BET} and transitivity, we find $p \sim p_1$.
    \end{proof}
\end{alemma}

For distinct $p,q$, $\aff(p,q) = \left.\left\{ q + t \cdot (p-q) \right| t \in \R \right\}$ is the straight line trough $p$ and $q$. A strict preference between $p,q$ sets all lotteries on their line in relation.
\begin{alemma}\label{Lemma: strict preference line segments}
    Let $\succeq$ fulfill \ref{Axiom: BET} and let $p,q \in \L$ be with $p \succ q$ and set $p_t := q + t \cdot (p-q) \in \aff(p,q) \cap \L$. Then, $p_0 = q, p_1 = p$ and 
    \begin{equation}
        \begin{cases}
            q \succ p_t &, \tif t <0,\\
            p \succ p_t \succ q &, \tif 0<t<1,\\
            p_t \succ p &, \tif 1<t.
        \end{cases}
    \end{equation}
    \begin{proof}[Proof of Lemma \ref{Lemma: strict preference line segments}]
        The cases for $t=0,1$ are clear. \emph{Case $t \in (0,1)$}: Since $p \succ q$ we have $p \succ p_t = t p + (1-t) q \succ q$ by \ref{Axiom: BET}. \emph{Case $1 < t$}: Note that $p = \alpha p_t + (1-\alpha) q$ for $\alpha := \tfrac{1}{t}$. Assuming that $q \succeq p_t$ leads to the contradiction $q \succeq p \succeq p_t$ by \ref{Axiom: BET}. Thus, $p_t \succ q$ by completeness, and hence $p_t \succ p \succ q$ by \ref{Axiom: BET}. \emph{Case $t <0$}: Now, $q=\alpha p_t + (1-\alpha) p$ for $\alpha := \frac{1}{1-t}$. Assuming that $p_t \succeq p$ leads to the contradiction $p_t \succeq q \succeq p$ by \ref{Axiom: BET}. Thus, $p \succ p_t$ by completeness, and hence $p \succ q \succ p_t$ by \ref{Axiom: BET}.
    \end{proof}
\end{alemma}

Under \ref{Axiom: IA}, translations of indifference sets remain indifference sets.
\begin{alemma}\label{Lemma: indifference translations}
    Let $p, q \in \L$. Then $\left.\left\{ r + (q-p) \right| r \in \L_{\sim p} \right\} \cap \L \subseteq \L_{\sim q}$.
    \begin{proof}[Proof of Lemma \ref{Lemma: indifference translations}]
    Note that $q = p+(q-p) \in \left.\left\{ r + (q-p) \right| r \in \L_{\sim p} \right\} \cap \L$. If this set is a singleton, there is nothing to prove. Let thus $q \neq q' = p' + (q-p) \in \L$ for some $p' \in \L_{\sim p}$. Define $z := \tfrac{1}{2} p + \tfrac{1}{2}q' = \tfrac{1}{2} p'+ \tfrac{1}{2} q$. Since $p \sim p'$, we have $z \sim \tfrac{1}{2} p + \tfrac{1}{2} q$ and thus $q' \sim q$ by \ref{Axiom: IA}.
    \end{proof}
\end{alemma}

We now proof Theorem \ref{Theorem: main theorem}. 
    \begin{proof}[Proof of Theorem \ref{Theorem: main theorem}]
        It is standard that \ref{Axiom: IA} and \ref{Axiom: MC} imply an expected utility representation, cf.\ \cite[Proposition 6.B.3]{mas1995microeconomic}. It thus suffices to prove that \ref{Axiom: IP} and \ref{Axiom: MC} are equivalent under \ref{Axiom: IA}. Proposition \ref{Proposition: Mixture Continuity implies IP} establishes that \ref{Axiom: MC} implies \ref{Axiom: IP} even in the absence of \ref{Axiom: IA}. Consequently, it only remains to show that  weak \ref{Axiom: IP} implies \ref{Axiom: MC} under \ref{Axiom: IA}.\\
        To this end, let $p_1, \ldots, p_{n} \in \L_{\sim p_1}$ such that $\{ p_k - p_1 \}_{k=2}^{n}$ are linearly independent in $\R^{n}$. Then $A := \aff(p_1, \ldots, p_{n})$ is a hyperplane of $\R^{n}$, i.e., $\dim A = n-1$. Hence, there exists a normal vector $\v \in\R^{n}$ such that $A = \left.\left\{ q \in \R^{n} \right| \langle q-p_1,\v \rangle = 0 \right\}$, where $\langle \cdot,\cdot \rangle$ is the scalar product on $\R^{n}$. Consider now any $r \in \L$. I will show an even stronger form of continuity, namely that $\L_{\succ r}$, and analogously $\L_{\prec r}$, are open. \ref{Axiom: MC} then follows since $[0,1] \to \R^n, \alpha \mapsto \alpha p + (1-\alpha)q$ is continuous (for all $p,q$) and $\L_{\preceq r} = \L \setminus \L_{\succ r}$ is closed. To see why $\L_{\succ r}$ is open, consider the hyperplane parallel to $A$ that passes through $r$ and partitions $\R^{n}$ (and thus $\L$) into three convex parts: $H^+_r := \left.\left\{ q \in \R^{n} \right| \langle q-r,\v \rangle > 0 \right\}$, $H_r := \left.\left\{ q \in \R^{n} \right| \langle q-r,\v \rangle = 0 \right\}$ and $H^-_r := \left.\left\{ q \in \R^{n} \right| \langle q-r,\v \rangle < 0 \right\}$. If $\L_{\succ r} = \emptyset$, there is nothing to prove. Assume thus that there is a $r^* \in \L_{\succ r}$. Note that $H_r = \{ q + (p_1-r) \mid q \in A \}$ and $A \cap \L \subseteq \L_{\sim p_1}$ by Lemma \ref{Lemma: affine space preserves indifference}. By Lemma \ref{Lemma: indifference translations}, $H_r \cap \L \subseteq \L_{\sim r}$. I can assume $r^* \in H^+_r$ (otherwise consider $-\v$). I claim $\L_{\succ r} = H^+_r \cap \L$, which is an open set in $\L$.

        \noindent%
        \emph{``$\subseteq$''}: Let $q \in \L_{\succ r}$ and assume by means of contradiction $q \in H^-_r$. Then, choosing $\alpha := \frac{-\langle q-r,\v \rangle}{\langle r^*-r,\v\rangle - \langle q-r, \v \rangle} \in (0,1)$ we have $\alpha r^* + (1-\alpha)q \in H_r \cap \L \subseteq \L_{\sim r}$, which violates $\alpha r^* + (1-\alpha)q \succ  \alpha r + (1-\alpha) q \succ r$, which we obtain from applying \ref{Axiom: IA} to $r^*,q \succ r$.

        \noindent%
        \emph{``$\supseteq$''}: Let $q \in H^+_r \cap \L$. Consider the hyperplanes $H_q, H_{r^*}$ parallel to $H_{r}$ and passing through $q$ resp.\ $r^*$. If $\langle r^*-r,\v \rangle\geq\langle q-r,\v\rangle$, $q$ lies between $H_{r^*}$ and $H_r$. Similar to above, we thus find an $\alpha \in [0,1)$ with $\alpha r + (1-\alpha) r^* \in H_{q} \cap \L$, which is indifferent to $q$ by Lemma \ref{Lemma: indifference translations}. By Lemma \ref{Lemma: strict preference line segments}, we thus have $r^* \succeq q \succ r$. Finally, if $\langle q-r,\v\rangle > \langle r^*-r,\v \rangle$, $r^*$ lies between $H_r$ and $H_q$. Connecting $r$ and $q$, we find an $\alpha \in (0,1)$ with $ \alpha r + (1-\alpha) q \in H_{r^*}\cap \L \subseteq \L_{\sim r^*}$. Applying Lemma \ref{Lemma: strict preference line segments}, we find $q \succ r^* \succ r$. 
    \end{proof}
\end{document}